\documentclass[journal,dvipsnames]{IEEEtran}

\usepackage{amsmath,amssymb,amsfonts}
\usepackage{array}
\usepackage{cite}
\usepackage{color}
\usepackage[pdftex]{graphicx}
\usepackage{nccmath}
\usepackage[percent]{overpic}
\usepackage{pifont}
\usepackage[caption=false]{subfig}
\usepackage{tikz}
\usepackage{upgreek}
\usepackage{verbatim}
\usepackage{fancybox}
\usepackage{pgf}
\usepackage{siunitx}
\usepackage{float}
\usepackage{xcolor}
\usepackage{multirow}
\usepackage{placeins}
\usepackage{pifont}

\input epsf
\newcommand{\fulltoday}{\ifcase\month\or
    January\or February\or March\or April\or May\or June\or
    July\or August\or September\or October\or November\or December\fi
    \space\number\day\space \number\year}
\def\cw{\columnwidth}

\def\iic{I\textsuperscript{2}C}
\def\iiw{I\textsuperscript{2}We}
\def\Vrx{V_{\mathrm{RX}}}
\def\Vhys{\Delta V_{\mathrm{hys}}}
\def\Rd{R_{\mathrm D}}
\def\Rtex{R_{\mathrm{tex}}}
\def\Rvar{R_{\mathrm{var}}}
\def\Lpara{L_{\parallel}}
\def\Ts{T_{\mathrm S}}
\def\Ltx{L_{\mathrm{TX}}}
\def\Lrx{L_{\mathrm{RX}}}
\def\uH{\si{\micro\henry}}
\def\nF{\si{\nano\farad}}
\def\us{\si{\micro\second}}
\def\kHz{\si{\kilo\hertz}}
\def\ohm{\si{\ohm}}
\def\mV{\si{\milli\volt}}
\def\Ctex{C_{\mathrm{tex}}}

\newcommand{\reffig}[1]{Fig.~\ref{fig:#1}}
\newcommand{\reftab}[1]{Table~\ref{tab:#1}}
\newcommand{\refeq}[1]{(\ref{eq:#1})}
\newcommand{\refsec}[1]{Section~\ref{sec:#1}}

\usepackage[hidelinks]{hyperref}

\ifCLASSINFOpdf
\else
\fi

\begin{document}
\title{UART for Wearables (U4We): DC Power and Carrierless Signal Transfer over Conductive Textiles}

\author{Akihito~Noda,~\IEEEmembership{Member,~IEEE}%
\thanks{A. Noda is with the School of Systems Engineering, Kochi University of Technology, Kochi, 782-8502 Japan.}%
\thanks{Accepted manuscript. Accepted for publication in \textit{IEEE Transactions on Circuits and Systems---II: Express Briefs}. DOI: \href{https://doi.org/10.1109/TCSII.2026.3735640}
{10.1109/TCSII.2026.3735640}}%
\thanks{This research was supported in part by JSPS KAKENHI JP25K03109 and JST CRONOS, Japan, Grant Number JPMJCS25N4.}}

\markboth{Accepted manuscript}%
{Noda: UART for Wearables (U4We)}

\maketitle

\begin{abstract}
This brief presents a circuit-level conductive-textile interconnection technique for batteryless distributed wearable modules. Two conductive textile layers separated by an insulating fabric layer are used as a garment-wide shared bus that simultaneously conveys DC power and pulse-based data signals without point-to-point wiring. To minimize the circuit overhead of each module, universal asynchronous receiver/transmitter (UART) pulses are directly coupled onto the textile through AC-coupling capacitors without carrier modulation. The textile is modeled as a lumped resistance--capacitance network. The textile waveform is then modeled as the transient response of a second-order circuit. From this model, design conditions for comparator-based waveform recovery and high-bitrate transmission are analytically derived. The resulting design framework determines whether a given combination of data rate, textile capacitance, and resistance is feasible, and also provides the corresponding design range of the decoupling inductors. The required textile parameters are characterized, and robustness to parameter drift is analyzed. These results establish a basic methodology for textile-based simultaneous power and data transfer.
\end{abstract}

\begin{IEEEkeywords}
E-textiles, power and data transfer, wearable communication.
\end{IEEEkeywords}

\IEEEpeerreviewmaketitle

\bstctlcite{IEEEexample:BSTcontrol}

\section{Introduction}

\IEEEPARstart{W}{earable} electronics are increasingly used for biological monitoring and human--machine interaction. Emerging applications such as distributed sensing \cite{Wicaksono2020ETeCS,Lin2017SmartSock}, haptic garments \cite{furukawa2019synesthesia}, and illuminated clothing \cite{song2022water} require multiple electronic modules to be distributed over a garment surface. A key challenge is how to interconnect these modules while preserving garment flexibility and comfort.

Wireless links require per-node antennas and RF circuitry and do not distribute power, so each module needs a battery. RFID (radio frequency identification)/NFC (near-field communication) sensors can operate batterylessly \cite{Lin2020NearFieldClothing,Zou2025NFCRFIDReview} but are sensitive to reader coupling.
Human-body communication \cite{Zhao2017HBCReview} likewise lacks a power-supply mechanism for multiple nodes. Batteryless garment-scale interconnection thus remains challenging.

Conductive textiles provide large-area interconnection while retaining fabric flexibility, e.g., patterned \cite{Stoppa2014CriticalReview,Tseghai2020Overview}, printed \cite{Paul2014ScreenPrinted,DEJENE2025100629}, and embroidered \cite{Weder2015EmbroideredElectrode,Lin2022LiquidMetalEmbroidery} wiring. However, such predefined routing precludes module relocation. A more flexible approach uses two conductive textile layers separated by an insulating layer as a garment-wide shared medium to which modules can be attached at arbitrary positions.

Existing methods for simultaneous power and data transfer over conductive textiles can be broadly categorized by their multiplexing scheme. Time-division multiplexing (TDM) alternates power delivery and data transmission over time~\cite{textilenet,zhu2020fast}. Frequency-division multiplexing (FDM) superimposes data on the DC power line using high-frequency carriers~\cite{nodaTBCAS2019}. FDM is attractive for power-consuming applications because it enables continuous power delivery. However, it typically requires carrier generation and modulation/demodulation circuits, which increase the complexity of each wearable module. 

In this context, this work proposes UART for wearables (U4We), a carrierless UART (universal asynchronous receiver/transmitter) pulse transmission scheme for conductive textile transmission lines that enables simultaneous power and data transfer with minimal circuit complexity. The proposed scheme targets applications such as garment-integrated LED arrays~\cite{nodaIMS2017}, distributed motion/physiological sensor nodes, and haptic actuator arrays~\cite{yoshimotoCCNC2026}, in which tens of small modules must be powered and individually addressed over a garment (\reffig{concept}). In such systems, the key requirement is a low-overhead shared medium rather than high throughput. 

The main contributions are as follows: 1) an AC-coupled pulse transmission scheme using a standard UART interface without carrier generation or demodulation; 2) an analytical circuit model that describes the transient response and derives the critical-damping design condition; and 3) an experimental demonstration of UART communication up to 1~Mb/s using a simple interface circuit. UART is used because it is available in most microcontrollers (MCUs). The concept applies to other pulse-based signaling as well. A preliminary demo abstract~\cite{noda2025ccnc} is extended here with the circuit model, design criteria, and validation. Sections~II--V then present the system concept, circuit model, design criteria, and experimental validation in this order.

\section{Simultaneous DC Power and UART Transfer}
\label{sec:dcuart}

\subsection{Overview}
\begin{figure}[t]
	\centering
    \setlength{\abovecaptionskip}{0pt}
\begin{overpic}[width=0.9\cw]{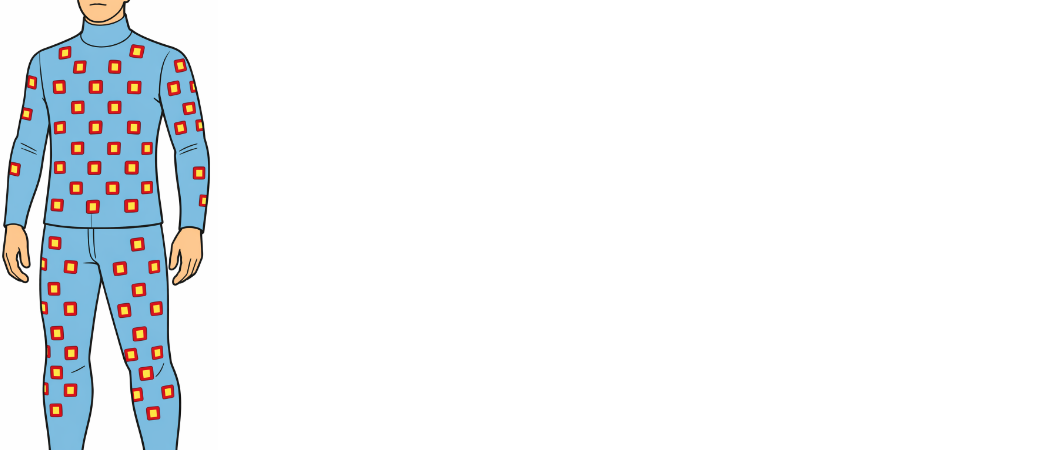}
    \put(0,0){\includegraphics[width=0.9\cw]{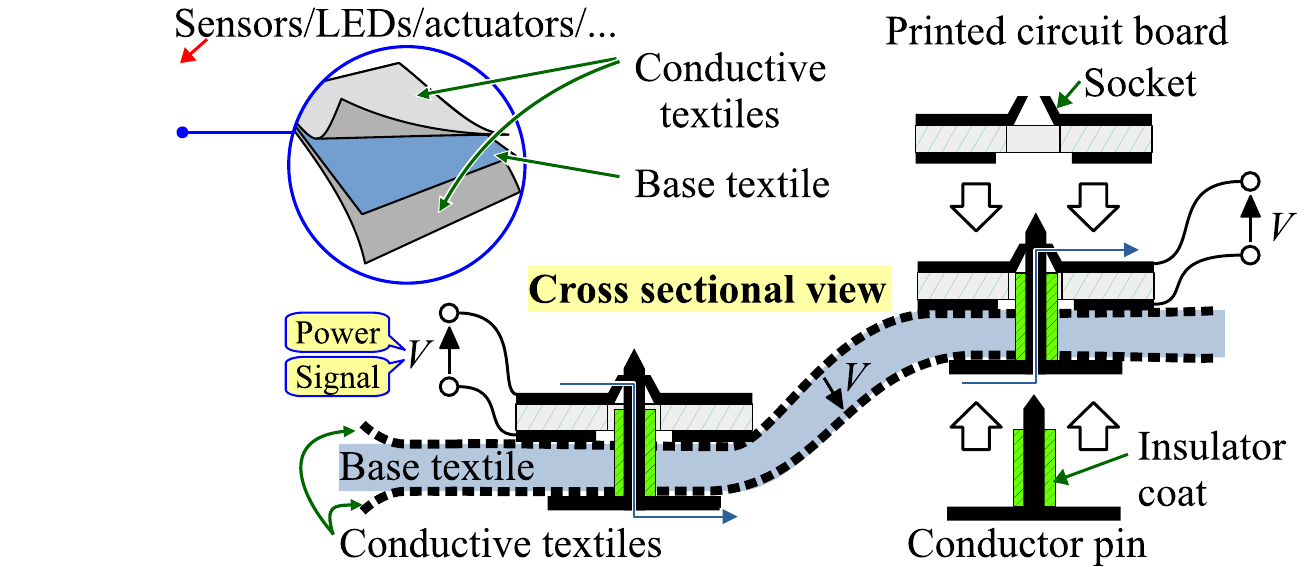}}
\end{overpic}	\caption{Concept: wearable modules attached to double-sided conductive textile via pin-and-socket fasteners share DC power and digital pulses.}
	\label{fig:concept}
\end{figure}

As shown in \reffig{concept}, the garment consists of two conductive textile layers separated by an insulating layer, and modules are attached by a pin-and-clutch fastener that contacts both layers without sewing or adhesives. The textile thus functions as a two-dimensional shared medium that distributes power and signals to modules at arbitrary positions.
The skin-side conductive layer is at a single common potential over its entire surface, whichever rail it is assigned to, and the bus voltage appears only across the thin insulating layer between the two conductive layers. Unlike human-body communication systems, the proposed method does not rely on signals induced in the human body.

\subsection{Broadcast Operation and Medium Access}
Signals injected at one point propagate over the entire medium, and the physical layer therefore forms a broadcast bus. Although UART is normally used for point-to-point links, the proposed system broadcasts the UART waveform over the shared conductive textile bus.
Node selection is then handled at the protocol level using framed data with destination addresses. Each module therefore processes only the frames addressed to it.

Simultaneous transmission by two nodes corrupts the superimposed frames, as in any multi-drop bus such as \iic{} or RS-485. Collision avoidance belongs to the medium-access-control layer, whereas this brief addresses the physical layer. Here, master-initiated half-duplex access is adopted. An RX module drives the bus only when polled, collisions are detected by UART framing/checksum errors, and listen-before-talk can be added in firmware.

\subsection{Physical-Layer Considerations}
The proposed scheme confines the signal band so that the entire bus is far smaller than the signal wavelength; the next section quantifies this lumped-element regime.

Effects that dominate in high-frequency systems are therefore negligible by design: characteristic impedance and reflections, parasitic transmission-line modes, stray capacitance between the textile and the body, and module-to-module electromagnetic coupling or crosstalk. All modules observe practically the same waveform, with propagation delay negligible relative to the symbol duration. Slow baseline disturbances from layer-spacing variation, contact variation, and body-induced interference are tracked by the per-module adaptive threshold at each receiver.

\begin{figure}[t]
	\centering
    \setlength{\abovecaptionskip}{0pt}
	\includegraphics[width=0.9\cw,trim=0 2mm 0 0,clip]{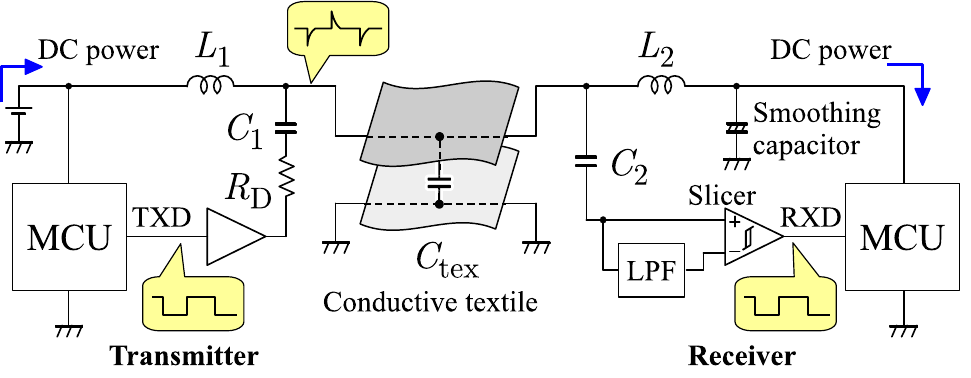}
	\caption{Simultaneous DC power and serial data transmission on textile.
    }
	\label{fig:uart}
\end{figure}

DC power and UART pulses are multiplexed with an inductor and a capacitor (\reffig{uart}). Power is supplied through the inductor, which isolates the power path from signal components, while the UART signal couples through a capacitor that blocks DC. Continuous power delivery and pulse signaling thus coexist without disturbing the DC operating point.
The smoothing and noise-bypass capacitors belong to the module side of these signal-band decoupling inductors. A DC source or capacitor must not be attached to the textile directly, since $\Ctex$ is a fundamental design parameter of the proposed method.

The interface circuit must (i)~drive a sufficiently large transient voltage onto the DC-biased, highly capacitive textile and (ii)~recover the digital waveform from the resulting spike-shaped transient response, rather than from a rectangular pulse waveform. The next section models this transient to establish the design methodology.

\section{Circuit Model and Response Waveform}
\label{sec:model}

\begin{figure}[t]
	\centering
    \setlength{\abovecaptionskip}{2pt}
    \includegraphics[width=0.9\cw]{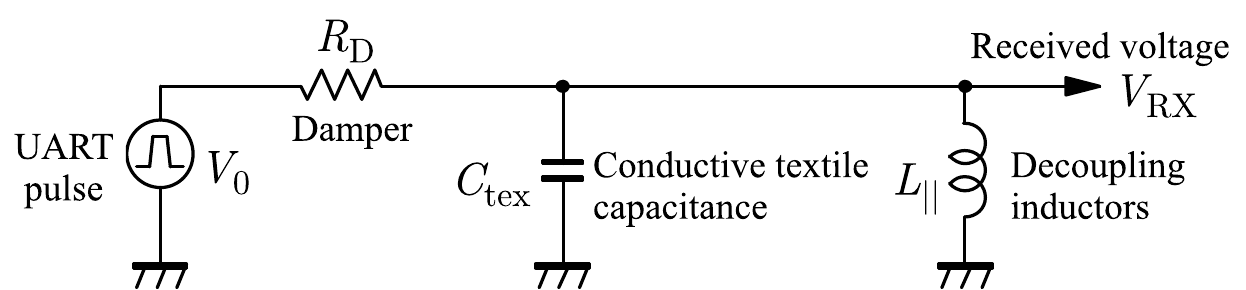}
	\caption{AC equivalent circuit of the TX--textile--RX signal path after open/short approximations at the signal frequency, with all module inductors combined into $\Lpara$.}
	\label{fig:equivcircuit}
\end{figure}

The AC equivalent circuit of the signal path through the conductive textile is shown in \reffig{equivcircuit}. It is derived by applying the following approximations.

First of all, the textile shared bus is modeled as a single lumped capacitance $\Ctex$. This is justified because the garment is electrically short at the occupied bandwidth. The Fourier magnitude of the critically damped spike \refeq{tran} is proportional to $\omega_0/(\omega^2+\omega_0^2)$, rolling off at $-40$~dB/decade above $\omega_0$, and more than 99\% of the spike energy lies below $3.37\omega_0/2\pi$. With $\omega_0/2\pi$ of a few megahertz for 1~Mbaud communication, the occupied bandwidth is less than 10~MHz, corresponding to a free-space wavelength longer than 30~m. For wearable applications, the garment size is assumed to be far smaller than this wavelength. Much higher bit rates and/or much larger textile dimensions would require a distributed RLGC model, which is beyond the scope of this brief.

At the signal frequency, the TX DC source and the AC-coupling capacitor are treated as short circuits. On the RX side, the data-slicer input is open and the smoothing capacitor following the decoupling inductor is a short, leaving only the inductor. The remaining significant elements are the textile capacitance $\Ctex$, the TX/RX inductors, which are combined in parallel into the equivalent inductance $\Lpara$, and the damping resistor, as shown in \reffig{equivcircuit}.

Based on this simplified model, the transient response of the received pulse waveform can be derived analytically. The received voltage is expressed in the Laplace domain as $\Vrx(s) = V_0(s)\,(1/\Rd)\big/\!\left(1/\Rd + s\Ctex + 1/s\Lpara\right)$, where $\Lpara:=L_1\parallel L_2 \parallel \cdots \parallel L_N$, and $N$ is the total number of TX/RX modules connected to the textile bus. To analyze the transient response at a pulse edge, a unit-step input $V_0=1/s$ is assumed, yielding

\begin{equation}
    \Vrx(s)
    =
    \dfrac{2\zeta}{\omega_0}
   \dfrac{\omega_0^2}{s^2+2\zeta\omega_0 s+ \omega_0^2},
    \label{eq:Vrx}
\end{equation}
where
\begin{equation}
    \omega_0 := 1 \left/ \sqrt{\Lpara\Ctex} \right., \qquad
    \zeta := \dfrac{1}{2\Rd} \sqrt{\dfrac \Lpara \Ctex}. \label{eq:omega}
\end{equation}

\begin{figure}[t]
    \centering
    \setlength{\abovecaptionskip}{0pt}
    \scriptsize
    \includegraphics[width=8.5cm]{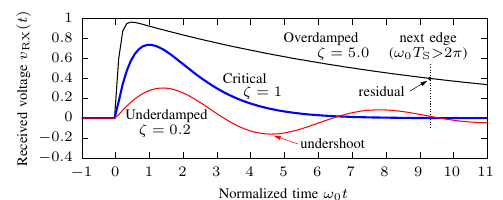}
    \caption{Step response waveforms. The undershoot for $\zeta<1$ and the residual remaining at the next pulse edge for $\zeta>1$ are annotated. The next-edge line indicates an example timing of the next pulse edge satisfying the design condition $\omega_0\Ts>2\pi$.}
	\label{fig:stepresponse}
\end{figure}

Equation~\refeq{Vrx} is the standard second-order response, where $\zeta$ is the damping ratio. For $\zeta<1$, the response becomes oscillatory, which is unsuitable for reliable recovery using a simple data slicer. In contrast, for $\zeta>1$, the response becomes overdamped and converges too slowly. Therefore, the critically damped case, $\zeta=1$, is of primary interest in the proposed receiver. Representative waveforms are shown in \reffig{stepresponse}.

For $\zeta=1$, \refeq{Vrx} becomes $\Vrx(s)=2\omega_0/(s+\omega_0)^2$ and its inverse Laplace transform gives
\begin{equation}
    v_{\rm RX}(t)
    \equiv
    \mathcal{L}^{-1}\{V_{\rm RX}(s)\}
    =
    2\omega_0 t {\mathrm e}^{-\omega_0 t}.
    \label{eq:tran}
\end{equation}

Under critical damping (\reffig{stepresponse}), each pulse edge is converted into a non-oscillatory spike of corresponding polarity that returns to zero without polarity reversal, so a comparator with minimal hysteresis can recover the waveform. The spike peaks at $\omega_0 t=1$ and decays substantially by $\omega_0 t\approx2\pi$. Hence, if the pulse length $\Ts$ satisfies $\omega_0 \Ts>2\pi$, successive edges do not interfere. Critical damping and sufficiently fast decay are thus the two receiver requirements, leading directly to the design conditions in the next section.

\section{Design Criteria}
\label{sec:criteria}

The transient response is governed by $\Ctex$, $\Lpara$, and $\Rd$. $\Ctex$ is set by the textile dimensions and materials and is treated as given. The design parameters are $\Lpara$ and $\Rd$, subject to $\Rd > \Rtex$, where $\Rtex$ denotes the total series resistance of the signal path, including textile spreading and fastener contact resistance.

Garment-scale $\Ctex$ is typically on the order of nanofarads, and $\Rtex$ of metal-plated fabrics on the order of $1\,\Omega$, which sets the practical design range.

The first requirement is critical damping, $\zeta=1$ (\refsec{model}). Substituting it into \refeq{omega} gives
\begin{equation}
    2\Rd = \sqrt{\Lpara/\Ctex}.
    \label{eq:Rd_crit}
\end{equation}

The second is sufficiently fast decay, $\omega_0 \Ts > 2\pi$ (\refsec{model}), which with \refeq{omega} yields the upper bound
\begin{equation}
    \Lpara < \dfrac{\Ts^2}{4\pi^2 \Ctex}.
    \label{eq:Lpara_upper}
\end{equation}

A lower bound on $\Lpara$ is obtained from the critical-damping relation \refeq{Rd_crit} together with the constraint $\Rd > \Rtex$ as
\begin{equation}
    \Lpara > 4\Rtex^2 \Ctex.
    \label{eq:Lpara_lower}
\end{equation}

\begin{figure}[t]
\centering
\setlength{\abovecaptionskip}{0pt}
    \scriptsize
\includegraphics[width=8.4cm]{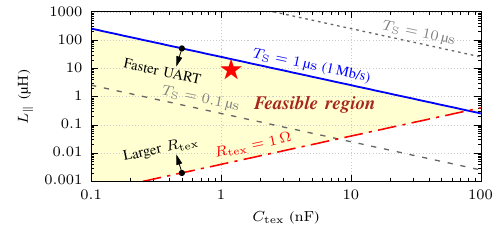}

\caption{Design space of $\Lpara$ vs.\ $\Ctex$. $\Ts$ sets the upper bound and $\Rtex$ (critical damping) the lower bound. Highlighted: $\Ts=1\,\us$, $\Rtex=1\,\Omega$. The region shrinks for shorter $\Ts$ or larger $\Rtex$. Star: design example of \refsec{implementation} ($\Ctex=1.2\,\nF$, $\Lpara=8.2\,\uH$).}
\label{fig:designspace}

\end{figure}

The design constraints are summarized in \reffig{designspace}, where $\Ts$ sets the upper bound of the feasible $\Lpara$ region and $\Rtex$ sets the lower bound.

In practice, $\Ctex$ and $\Rtex$ deviate from their nominal values with moisture, pressure, and geometry, and the same module circuit may be attached to garments of different sizes. Deviations from $\zeta=1$ produce a finite undershoot ($\zeta<1$) or a residual at the next edge ($\zeta>1$) in the received waveform (\reffig{stepresponse}). An undershoot of magnitude $V_{\mathrm{us}}>\Vhys$ falsely triggers the comparator. The hysteresis width must therefore satisfy $V_{\mathrm{peak}}>\Vhys>V_{\mathrm{us}}$. Because $V_{\mathrm{peak}}\gg\Vhys$ by design, the tolerance is asymmetric, and the undershoot side is the binding constraint. The margin is still wide because the deviations grow slowly; $V_{\mathrm{us}}\approx0.01V_{\mathrm{peak}}$ at $\zeta=0.82$. Larger or accumulated drifts are compensated by readjusting the trimmer, which restores $\zeta=1$ through $\Rd=\frac12\sqrt{\Lpara/\Ctex}$.

\section{Hardware Implementation and Experimental Validation}
\label{sec:implementation}

The prototype validates the model in a one-way configuration with one TX commanding multiple RX modules, representative of an LED array. Bidirectional transfer can use TX/RX role switching under the half-duplex, master-initiated access described in \refsec{dcuart}.

For this configuration, a practical choice is $\Ltx$ near the upper bound of $\Lpara$ and $\Lrx\gg\Lpara$. With fixed inductors, $\Lpara=\Ltx\parallel(\Lrx/M)$, so its dependence on $M$ is mitigated.

The representative design example is summarized in \reftab{design_example}, and the case of $M=1$ is indicated by the star in \reffig{designspace}. The table also lists the measured bus parameters: the sheet resistance \cite{MACfab}, and the capacitance per unit area obtained from the lumped capacitance measured between the assembled layers with a digital multimeter. As a practical example including fastener contact and sheet spreading resistances, the loop resistance from a feeding fastener through one layer to a layer-shorting fastener and back through the other was 0.37~$\Omega$ at a 30~cm separation. This loop is exactly the series resistance inserted in the TX--RX path. $\Ctex$ and $\Rtex$ are effective parameters of the assembled structure. For another garment, they should be measured likewise, i.e., $\Ctex$ between the layers and $\Rtex$ between the attachment points, before selecting $\Lpara$ and $\Rd$ from \refsec{criteria} and fine-tuning the trimmer. All values are for flat, static, and dry conditions (\refsec{conclusion}). An in-line trimmer $\Rvar$ in the UART path sets the critical-damping condition. It is tuned once at setup while the textile waveform is observed on an oscilloscope. In the measurements below, $\Rvar$ was adjusted to reproduce underdamped, near-critical, and overdamped conditions.

\begin{table}[t]
\centering
\caption{Design Parameters and Characterized Textile Properties of the Prototype}
\label{tab:design_example}
\renewcommand{\arraystretch}{1.00}
\begin{tabular}{lcc}
\hline
\textbf{Symbol} & \textbf{Value} & \textbf{Unit} \\
\hline
$\Ctex$ & 1.2 & \nF \\
$\Rd$ & $\Rtex + R_{\mathrm{var}}$ &  \\
$\Rtex$ & 0.5 & $\Omega$ \\
$R_{\mathrm{var}}$ & \numrange{0}{200} & \ohm \\
$L_{\mathrm{TX}}$ & 8.3 & \uH \\
$L_{\mathrm{RX}}$ & 1000 & \uH \\
$\Lpara= \Ltx \parallel (\Lrx/M)$
 & ($M=1$) 8.2 &  \multirow{2}{*}{\uH} \\
\hspace{3mm}$M$: \# of RXs  & ($M=28$) 6.7 & \\
\hline
Sheet resistance (datasheet \cite{MACfab}) & $\le$0.1 & $\Omega/\square$ \\
Capacitance per area & 0.75 & pF/cm$^2$ \\
Insulator thickness & 0.6 & mm \\
Loop resistance, 10/20/30\,cm (4-wire) & 0.25/0.32/0.37 & $\Omega$ \\
\hline
$L_{\mathrm{RX}}$ saturation-current rating & 25 & mA \\
Rated aggregate RX current ($M=28$) & 0.70 & A \\
Worst-case common-path drop (0.70 A) & 0.50 & V \\
$L_{\mathrm{RX}}$ drop at 25 mA & 0.60 & V \\
Available RX voltage at 25 mA & 3.90 & V \\
Available RX power at 25 mA & 97 & mW \\
Available aggregate RX power ($M=28$) & 2.7 & W \\
\hline
\end{tabular}
\end{table}

\begin{figure}[t]
	\centering
    \setlength{\abovecaptionskip}{0pt}
	\includegraphics[width=0.94\cw]{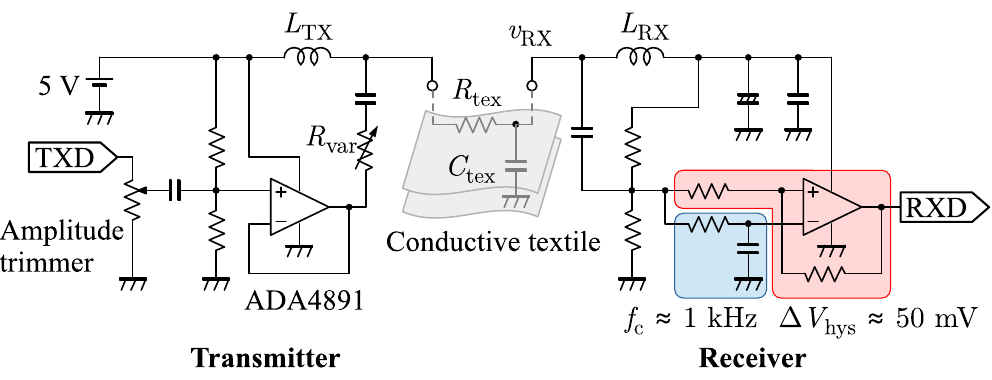}
	\caption{Schematic diagram of the experimental prototype (relevant portion).
    }
	\label{fig:prototype}
\end{figure}

\begin{figure}[t]
	\centering
    \setlength{\abovecaptionskip}{0pt}
	\input{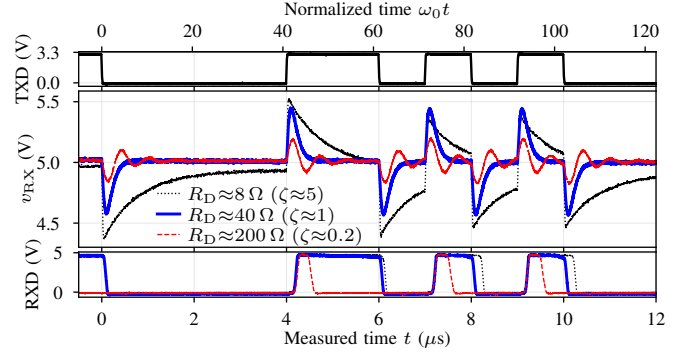}
	\caption{Measured waveforms at 1\,Mbaud ($\Ts=1\,\us$), single-RX configuration: (top) TX UART output (critical condition; common to all cases), (middle) textile-line voltage, (bottom) reconstructed RXD input, under underdamped, near-critical, and overdamped settings.}
	\label{fig:measuredwaveform}
\end{figure}

In the implementation (\reffig{prototype}), the positive and negative spikes are detected by a hysteresis comparator whose reference is generated by low-pass filtering the input, so that the threshold follows slow baseline variations. The 1~\kHz{} cutoff is chosen as a conservative separation frequency rather than as a measured upper bound on garment-induced disturbance. A recent review of soft bioelectronics reports that mechanical motion artifacts caused by body movements and physiological activities typically occur below 10~Hz~\cite{Yin2024MotionArtefact}. Thus, the 1~\kHz{} cutoff is well above frequencies commonly associated with body-motion-induced mechanical artifacts, while remaining more than three orders of magnitude below the characteristic transient frequency $\omega_0/2\pi\approx1.6$~MHz. The exact cutoff is therefore not critical. Baseline-wander spectra and dynamic $\Ctex$ and $\Rtex$ under controlled deformation were not characterized in this study. The hysteresis width $\Vhys \approx 50\,\mV$ is set according to the ordering $V_{\mathrm{peak}}>\Vhys>V_{\mathrm{us}}$ of \refsec{criteria} rather than as a unique optimum. It remains roughly an order of magnitude below the spike peak observed in \reffig{measuredwaveform}, preserving detection sensitivity, while the undershoot $V_{\mathrm{us}}$ caused by parameter drift exceeds the 50~\mV{} level only when $\zeta$ decreases to approximately 0.5, far beyond moderate drifts. Values in a similar range serve equally well.

The measurements in \reffig{measuredwaveform} were obtained for a representative single-RX configuration, probing the TX UART output, the RX pickup electrode, and the RX MCU UART input, for comparison with \reffig{stepresponse}.

The measured textile-line waveforms agree well with \reffig{stepresponse}. Under the near-critical condition the line returns to the 5~V baseline within each 1~\us{} symbol, making the peak amplitude independent of symbol history and yielding correct RXD recovery. The overdamped case leaves history-dependent peaks due to incomplete return, and the underdamped case produces oscillatory transients that cause false RXD transitions. These results validate the circuit model and the critically damped design strategy, and UART communication at 1\,Mb/s was confirmed under the representative condition.

The rating-based 28-RX DC power budget is summarized in \reftab{design_example} and is independent of 1~Mb/s signaling. Pulse edges are suppressed toward the local supply by the $\Lrx$--$C$ filter by $\omega_0^2\Lrx C\ge10^4$ ($C\ge0.1$~\si{\micro\farad}), to the sub-millivolt level. TX-supply ripple reaches the bus through $(\Lpara/\Ltx)\,\omega_0^2/(s^2+2\zeta\omega_0s+\omega_0^2)$; at $\zeta=1$ it has no peaking and gain at most $\Lpara/\Ltx\approx0.8$. Ripple above 1~\kHz{} should remain below $\Vhys$, while sub-\kHz{} components are tracked by the threshold. A load-step-induced voltage variation at an RX supply couples back to the bus through the same transfer function with $\Ltx$ replaced by $\Lrx$; at $\zeta=1$ its gain is at most $\Lpara/\Lrx\approx6.7\times10^{-3}$ for $M=28$.

\begin{figure}[t]
	\centering
    \setlength{\abovecaptionskip}{0pt}
\includegraphics[width=0.8\cw]{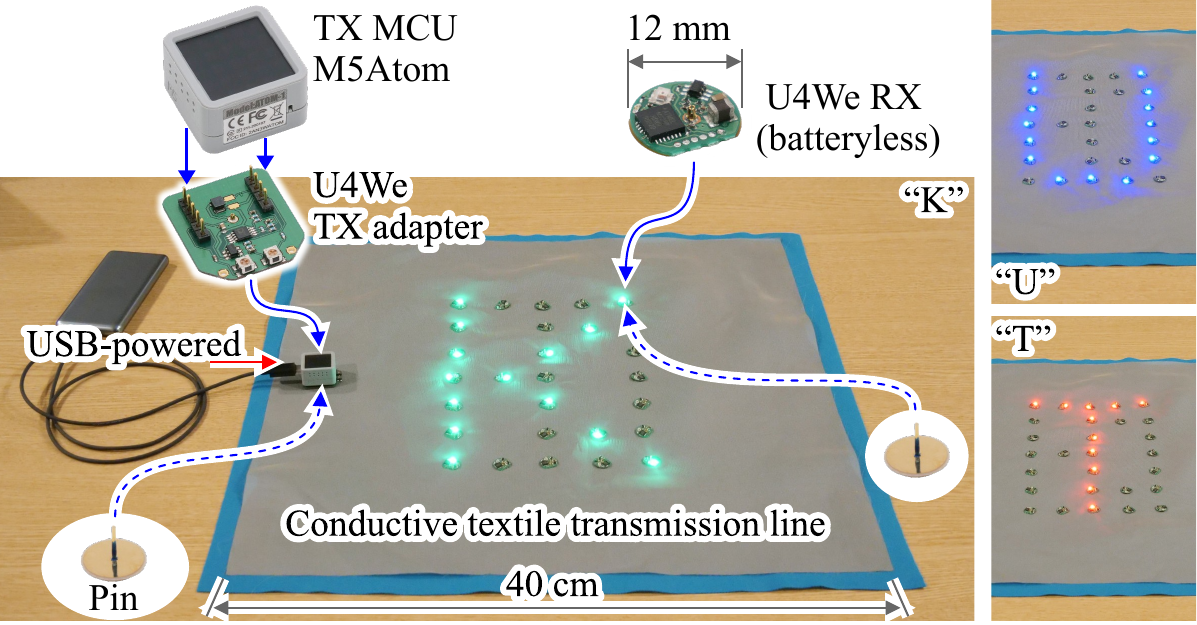}
\caption{28-RX prototype on the shared textile bus. The M5Atom TX with a U4We adapter selectively activates RX modules by address-based control.}
	\label{fig:28RXs}
\end{figure}

The 28-RX prototype (\reffig{28RXs}) has an M5Atom (M5Stack) TX with a U4We adapter and 28 custom 12~mm ATmega328P (Microchip) RX modules on 40~cm square Cu--Ni-plated fabric (DW-372N, MAC Corporation \cite{MACfab}) for addressed control.

Because $\Lrx \gg \Lpara$, increasing $M$ from 1 to 28 changes $\Lpara$ only from 8.2 to 6.7\,$\uH$ (critical $\Rd$ from 41 to 37\,$\ohm$). Accordingly, stable 28-RX operation was confirmed without readjusting the critical condition set in the single-RX case. The allowable module count is implementation-dependent rather than a fixed property of the bus: it is determined by whether the resulting $\Lpara(M)$, effective $\Ctex$, $\Rtex$, and $\Ts$ remain within the feasible region of \reffig{designspace}. Any non-negligible aggregate receiver parasitic capacitance is included in the effective $\Ctex$. At exact critical damping, \refeq{tran} has a peak amplitude of $2/{\mathrm e}$ of the applied step, which does not scale with $M$. The demonstrated $M=28$ is thus one experimentally verified operating point in this design space, not a universal upper limit. A supplementary video shows continuous LED-array operation while the textile test piece is lifted, shaken, and brought into contact with the human body, providing system-level evidence of robustness under these motions.

\reftab{comparison} compares U4We with textile TDM buses \cite{textilenet,zhu2020fast} and FDM-based \iiw{} \cite{nodaTBCAS2019}. U4We provides continuous, carrierless power/data transfer at 1~Mb/s; the listed values are those reported under differing operating conditions.

\begin{table}[t]
\centering
\caption{Comparison with Textile Power-and-Data Transfer Schemes}
\label{tab:comparison}
\renewcommand{\arraystretch}{1.15}
\setlength{\tabcolsep}{3pt}
\footnotesize
\begin{tabular*}{\cw}{@{\extracolsep{\fill}}lcccc@{}}
\hline
 & \cite{textilenet} & \cite{zhu2020fast} & \iiw{}~\cite{nodaTBCAS2019} & \textbf{U4We} \\
\hline
Scheme & TDM & TDM & FDM & Baseband \\
Cont.\ power & No & No & Yes & Yes \\
Carrierless & Yes & Yes & No & Yes \\
Interface$^{1}$ & \multicolumn{2}{c}{Charge/comm.\ switch} & ASK modem & LC + comp. \\
Data rate & 9.6\,kb/s & 1.5\,Mb/s$^{2}$ & 100\,kb/s & 1\,Mb/s \\
Nodes$^{3}$ & 5 & 10 & 8 & 28 \\
Size dep. & $RC$ limit & IR droop & Detuning & \refeq{Lpara_upper}, \refeq{Lpara_lower} \\
\hline
\end{tabular*}\\[2pt]
\parbox{\cw}{\scriptsize\raggedright
$^{1}$Per-node power/data multiplexing circuitry beyond the serial protocol itself (asynchronous serial for \cite{textilenet}, UART for \cite{zhu2020fast} and U4We, \iic{} for \cite{nodaTBCAS2019}). In \cite{textilenet}, \cite{zhu2020fast}, the power and data phases are alternated by switches (in \cite{zhu2020fast}, in the master, with an RS-485 transceiver per node).\\
$^{2}$Effective rate at 10 slaves (5~Mbaud burst UART with charging duty).\\
$^{3}$Numbers demonstrated in the respective papers, not theoretical upper limits.}
\end{table}

\section{Conclusions}
\label{sec:conclusion}

This work presented a conductive-textile interconnection scheme for batteryless, antennaless modules using a UART interface and a minimal added circuit, at 1\,Mb/s, near the typical upper rate of MCU UARTs. The main contribution of this work is a circuit model and design methodology for textile-based simultaneous power and data transfer, together with the effective-parameter characterization required to apply it to other textiles. Quantitative characterization of baseline and parameter variation under controlled deformation, stretching/washing durability, statistical contact characterization, bidirectional multi-node protocols, and distributed RLGC modeling for larger media or higher rates remain future work.

\FloatBarrier
\bibliographystyle{IEEEtran.bst}
\bibliography{mybib.bib}

\ifCLASSOPTIONcaptionsoff
  \newpage
\fi

\end{document}